\documentclass[aps,prb,reprint,groupedaddress]{revtex4-1}
\usepackage{graphicx}
\usepackage{dcolumn}
\usepackage{rotating}
\usepackage{amssymb}
\usepackage{mathptmx}
\usepackage{amsfonts}
\usepackage{amsmath}
\usepackage{bm}
\begin{document}
\title{Single crystal growth and study of the magnetic properties of the mixed spin-dimer system Ba$_{3-x}$Sr$_{x}$Cr$_{2}$O$_{8}$}
\author{Alsu Gazizulina$^1$, Diana L. Quintero-Castro$^{2,3}$, Andreas Schilling$^1$ }
%\email{alsu@physik.uzh.ch}
\affiliation{$^1$Physik-Institut, Universit\"{a}t Z\"{u}rich, Winterthurerstrasse 190, CH-8057 Z\"{u}rich, Switzerland \\
$^2$Helmholtz Zentrum Berlin f\"{u}r Materialien und Energie, D-14109 Berlin, Germany \\
$^3$Department of Mathematics and Natural Sciences, University of Stavanger, N-4036 Stavanger, Norway}
\date{\today}
\begin{abstract}
The compounds Sr$_{3}$Cr$_{2}$O$_{8}$ and Ba$_{3}$Cr$_{2}$O$_{8}$ are insulating dimerized antiferromagnets with Cr$^{5+}$ magnetic ions. These spin-$\frac{1}{2}$ ions form hexagonal bilayers with a strong intradimer antiferromagnetic interaction, that leads to a singlet ground state and gapped triplet states. We report on the effect on the magnetic properties of Sr$_{3}$Cr$_{2}$O$_{8}$ by introducing chemical disorder upon replacing Sr by Ba. Two single crystals of  Ba$_{3-x}$Sr$_{x}$Cr$_{2}$O$_{8}$ with $x=2.9$ (3.33\% of $mixing$) and $x=2.8$ (6.66\%) were grown in a four-mirror type optical floating-zone furnace. The magnetic properties on these compounds were studied by magnetization measurements. Inelastic neutron scattering measurements on Ba$_{0.1}$Sr$_{2.9}$Cr$_{2}$O$_{8}$ were performed in order to determine the interaction constants and the spin gap for $x=2.9$. The intradimer interaction constant is found to be  $J_0$=5.332(2) meV, about 4\% smaller than that of pure Sr$_{3}$Cr$_{2}$O$_{8}$, while the interdimer exchange interaction $J_e$ is smaller by 6.9\%. These results indicate a noticeable change in the magnetic properties by a random substitution effect.
\end{abstract} 
\maketitle	
\section{\label{Intr}Introduction}
Quantum magnetism is one of the most active areas of research in condensed matter physics. One of the reasons to study certain gapped quantum systems is the phenomenon of $equilibrium$ Bose-Einstein condensation (BEC) of magnetic quasiparticles, which is one of the most fascinating phenomena predicted by quantum mechanics [\onlinecite{zapf}]. The introduction of impurities in such systems has become an important topic since their discovery because quantum disordered spin systems behave differently from classical ones. Despite the difficulty in dealing with local quantum fluctuations, many important results have already been obtained in quantum disordered systems (see, e.g., Ref. [\onlinecite{dirty}] for a review).\par
In the isostructural antiferromagnetic Ba$_{3}$Cr$_{2}$O$_{8}$ and Sr$_{3}$Cr$_{2}$O$_{8}$ spin-dimer systems, the magnetic ions with a single electron in the 3$d$ shell (s = $\frac{1}{2}$) are located in an oxygen tetrahedron. The dominant antiferromagnetic exchange coupling $J_0$ between pairs of Cr$^{5+}$ ions creates spin dimers which are in a non-magnetic singlet state in a zero external magnetic field. By applying an external magnetic field larger than a critical value, so-called triplons can form in the network of dimers, which can be considered as a direct analogue to interacting bosons in a condensate [\onlinecite{growth_Ba, phases, growth_Sr, singh, chapon, aczel_sr, aczel_ba, kofu, diana_mag}]. The external magnetic field acts as a chemical potential in the limit of weak inter-dimer interactions [\onlinecite{gimr, nikuni}]. In real space, the BEC corresponds to a transition to an antiferromagnetic state with staggered magnetization. \par
This phenomenon has been observed, for example, in prototype systems such as BaCuSi$_{2}$O$_6$ [\onlinecite{jaime}] and TlCuCl$_{3}$ [\onlinecite{nikuni}]. However, in the case of exchange disorder, the nature of the ground state in a magnetic field and the critical behavior of the field-induced magnetic ordering are not sufficiently understood. Disorder can be induced in such systems, for example, by partial chemical substitution. The most prominent effect of such a chemically induced modification is a shift of the BEC phase boundary in the magnetic phase diagram [\onlinecite{dirty}]. However, experiments on non-magnetically substituted Ba$_{3}$(Cr$_{1-x}$V$_{x}$)O$_{8}$ suggest no significant change in the strength of the intradimer coupling $J_0$ and an increase of the effective interdimer coupling $J_e$ along with impurity substitution [\onlinecite{hong}]. Similar conclusions have been drawn on Sr$_{3}$Cr$_{2-x}$M$_{x}$O$_{8}$ where Cr was substituted by M = (V, Mn) [\onlinecite{chatto}]. Neither any strong suppression of the spin gap nor any signature of a change of the ordered magnetic state was observed up to a 10\% substitution level.\par
By introducing chemical disorder we may create a different critical behavior from the standard BEC one, namely the critical behavior of a Bose glass phase [\onlinecite{fisher}]. Investigations of possible Bose glass related transitions are very interesting although quite challenging. Such a transition was investigated in Tl$_{1-x}$K$_{x}$CuCl$_{3}$,  where a triplon localization in the doped system was observed [\onlinecite{yamada}]. Another doped system, Br-doped DTN, Ni(Cl$_{1-x}$Ba$_{x}$)$_{2}$-4SC(NH$_{2}$)$_{2}$, studies by nuclear magnetic resonance show a Bose-glass regime, where impurity states are strongly localized [\onlinecite{orlova}]. This system can be effectively described by two-level impurity states whose pairwise interaction is finite, although experimentally suppressed with distance. The doped system Ba$_{3-x}$Sr$_{x}$Cr$_{2}$O$_{8}$ to be discussed here represents another promising candidate for the observation of Bose-glass physics [\onlinecite{dirty}].\par
Recently, we reported a peculiar non-linear tuning of the magnetic intradimer interaction constant $J_0$ in a corresponding solid solution Ba$_{3-x}$Sr$_{x}$Cr$_{2}$O$_{8}$ by varying the Ba and Sr content [\onlinecite{grundmann_tuning}], as indirectly inferred from magnetization measurements on polycrystalline samples. However, there is no  corresponding information about respective data taken on single crystalline Ba$_{3-x}$Sr$_{x}$Cr$_{2}$O$_{8}$ samples up to now, nor has $J_0$ been directly measured, e.g., by inelastic neutron scattering techniques.\par
Here, we report on the growth of  Ba$_{3-x}$Sr$_{x}$Cr$_{2}$O$_{8}$ single crystals with $x=2.9$ (3.33\% mixing) and $x=2.8$ (6.66\% mixing) and the influence of chemical disorder on the magnetic and the structural properties. Moreover, we have performed inelastic neutron scattering measurements on single crystalline Ba$_{3-x}$Sr$_{x}$Cr$_{2}$O$_{8}$ with $x=2.9$ which show three excitation modes, corresponding to three twinned domains that confirm a monoclinic distortion. The hexagonal notation is kept throughout all the article. The interaction constants were determined from a random phase approximation (RPA) model, and the results are compared to those of Sr$_{3}$Cr$_{2}$O$_{8}$ and Ba$_{3}$Cr$_{2}$O$_{8}$, respectively.
\section{\label{sec:two}Experimental details and crystal growth}
\begin{figure}[htt]
\includegraphics[width=0.4\textwidth]{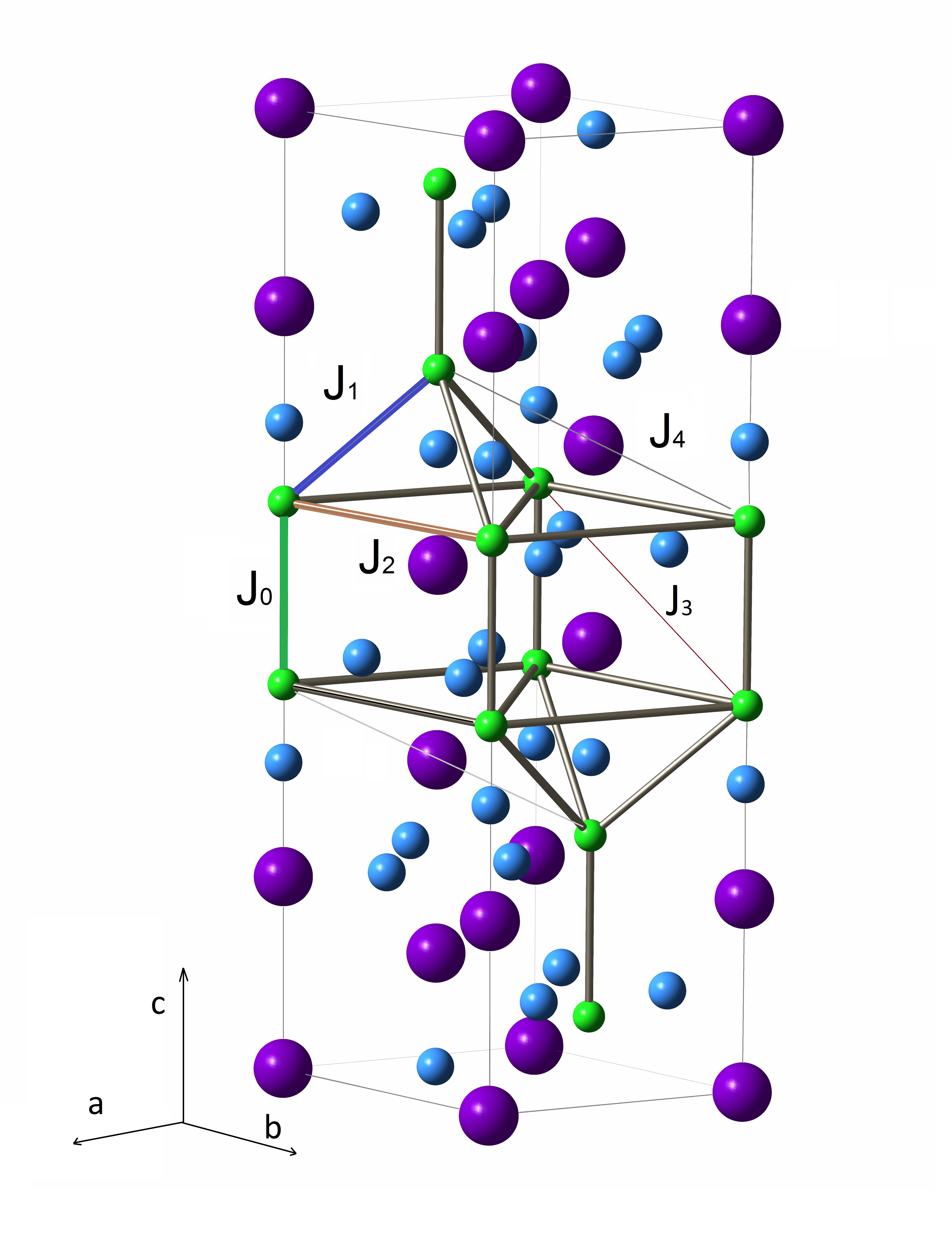}
\caption{Crystal structure of Ba$_{3-x}$Sr$_{x}$Cr$_{2}$O$_{8}$ at room temperature. Small green circles represent Cr ions, medium light blue circles are oxygen atoms and large violet circles stand for Sr or Ba ions}
\label{str}
\end{figure}

The crystal structure of Ba$_{3}$Cr$_{2}$O$_{8}$ presented in Figure \ref{str} is hexagonal with space group R$\bar{3}$m at room temperature. The nuclear structure of the solid solution Ba$_{3-x}$Sr$_{x}$Cr$_{2}$O$_{8}$ series has been shown to remain unchanged, but the cell parameters are linearly decreasing with an increase of the Sr content $x$ [\onlinecite{grundmann_influence, grundmann_structure}]. Single crystals of the unsubstituted end-members Sr$_{3}$Cr$_{2}$O$_{8}$ ($x=3$) and Ba$_{3}$Cr$_{2}$O$_{8}$ ($x=0$) can be grown using the traveling solvent floating zone method [\onlinecite{growth_Sr, growth_Ba}]. \par
In order to grow Ba$_{3-x}$Sr$_{x}$Cr$_{2}$O$_{8}$ crystals, high purity powders of SrCO$_{3}$  (Sigma-Aldrich 99.9\%), BaCO$_{3}$ (Sigma-Aldrich 99.98\%) and Cr$_{2}$O$_{8}$ (Sigma-Aldrich 99.9\%) were used as starting materials. Several samples were prepared with various Sr content $x$ using a solid-state reaction: \ $(3-x)$BaCO$_{3}$ + $x$SrCO$_{3}$ + Cr$_{2}$O$_{8}$ $\rightarrow$ Ba$_{3-x}$Sr$_{x}$Cr$_{2}$O$_{8}$ + 3CO$_{2}$. The powders were ground in a planetary ball mill (Pulverisette 5) and heated to 1250$^{\circ}$C for 24 hours, quenched in liquid nitrogen to avoid a reaction with atmospheric oxygen and avert impurity phases. This process was repeated three times, and the resulting powders were kept in vacuum. The synthesized powders were examined by powder X-ray diffraction at room temperature and found to be phase-pure with the correct hexagonal crystal symmetry  R$\bar{3}$m. The powders were then pressed at 2000 bar into cylindrical pellets. 
The single crystals were grown in the Crystal Laboratory at the Helmholtz Zentrum Berlin f\"{u}r Materialien und Energie (HZB) by using a high-temperature optical floating-zone furnace (Crystal Systems Inc. Model FZ-T-10000-H-VI-VPO) that was equipped with 300 W halogen lamps as a heat source and four ellipsoidal mirrors, thus achieving temperatures up to 2000$^{\circ}$C. Powder X-ray and Laue diffraction measurements were performed to check the structure of the powders, crystal purity and orientation. The relation between Sr and Ba contents was verified by EDX analysis at various points on the surface and on slices of the crystals in order to check for the crystal homogeneity.\par
Commonly, Ba$_{3}$Cr$_{2}$O$_{8}$ crystals are grown under argon atmosphere at a relatively high rate [\onlinecite{growth_Ba}], while Sr$_{3}$Cr$_{2}$O$_{8}$ single crystals are grown in flowing synthetic air [\onlinecite{growth_Sr}]. Sr$_{3}$Cr$_{2}$O$_{8}$ is stable at 1250$^{\circ}$C under pure oxygen atmosphere. However,  a transition of Sr$_{3}$Cr$_{2}$O$_{8}$ taking place around 775$^{\circ}$C  makes the single-crystal growth challenging due to a chemical reaction with oxygen that changes the oxidation state and phase. Oxidized Sr$_{3}$Cr$_{2}$O$_{8}$ is no longer stable in 1 atm oxygen atmosphere [\onlinecite{phases}]. Moreover, Sr$_{3}$Cr$_{2}$O$_{8}$ starts to react strongly with moisture and oxygen, and can decompose into SrCrO$_{4}$ and Sr$_{10}$Cr$_{6}$O$_{24}$(OH)$_{2}$ [\onlinecite{kisil}]. \par
We examined several preparation conditions for a successful crystal growth. We obtain a stable Ba$_{0.1}$Sr$_{2.9}$Cr$_{2}$O$_{8}$ single crystal by using the conditions similar to the preparation of Sr$_{3}$Cr$_{2}$O$_{8}$ crystals. However, with a subsequent increase of the Ba content $x$, the crystal growth becomes unstable, and the crystals tend to break after some time, which might be due to a chemical reaction upon cooling. The crystals of Ba$_{0.2}$Sr$_{2.8}$Cr$_{2}$O$_{8}$ grown under argon and 10\% oxygen in argon often have a rough surface and contain many crystal grains. This must be due to a different compositional phase diagram in argon atmosphere, and the necessary temperature to obtain a pure phase was probably not reached. Moreover, argon atmosphere may create oxygen deficiencies in the crystals. For the present experiment, we chose a single crystal immediately after its preparation to minimize the above mentioned problems. \par
\begin{figure}[htt]
\includegraphics[width=0.4\textwidth]{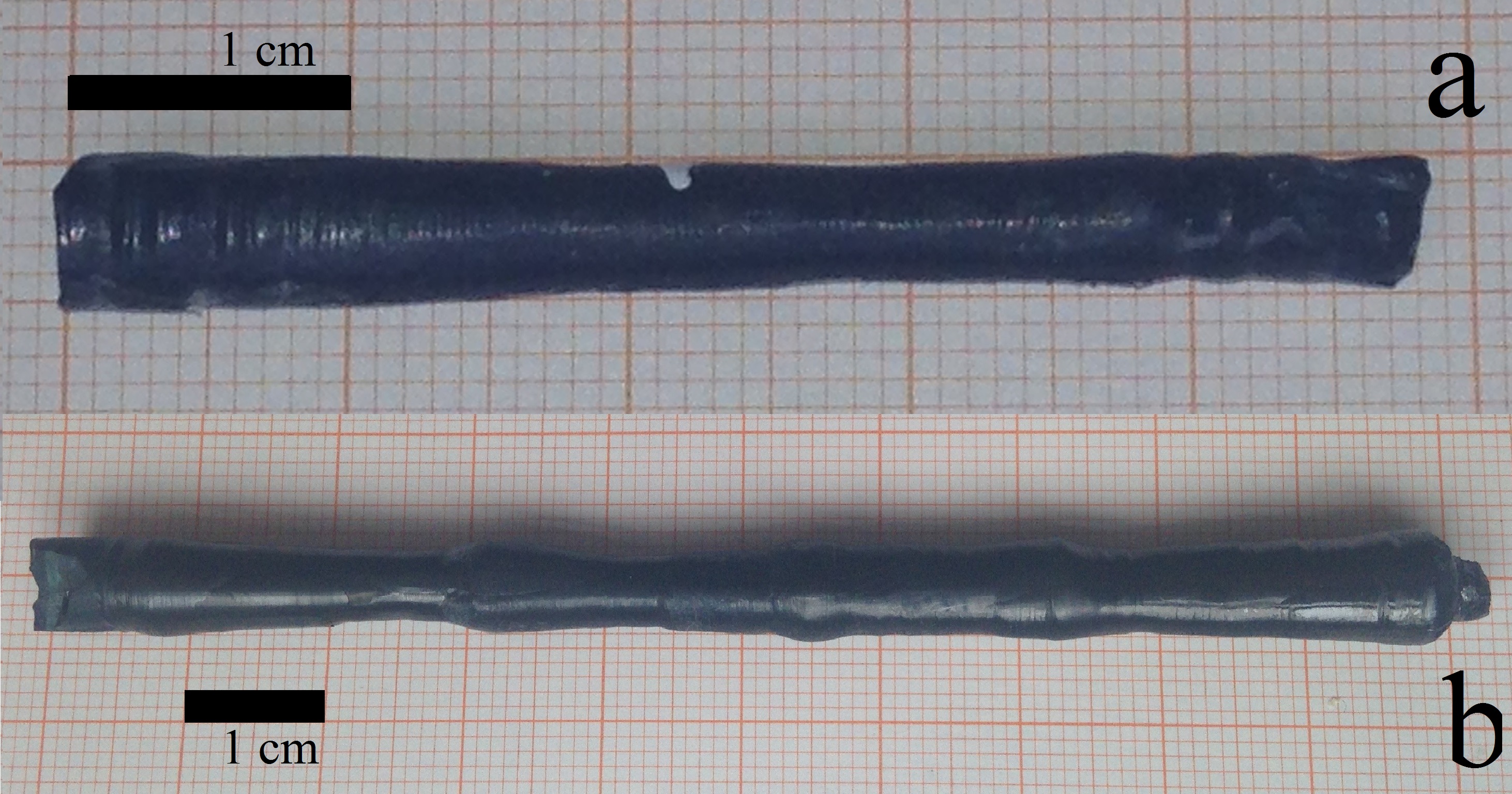}
\caption{As-grown samples of Ba$_{3-x}$Sr$_{x}$Cr$_{2}$O$_{8}$, (a) $x=2.9$, (b) $x=2.8$. X-ray Laue diffraction on  both samples show their crystalline nature, as single crystals or multiple crystallites with a shared orientation.}
\label{crystals}
\end{figure}

We used synthetic air flow (20.5\% oxygen in N$_{2}$) of 2 liters/min at ambient pressure for the $x=2.9$ sample. The seed and feed rods were counter-rotated at rates of 30 and 8 rpm, respectively, to obtain a homogeneously illuminated region of the sample. The seed rod was slowly moved out of the hot region for slow cooling and recrystallization. The growth rate is one of the conditions that was varied for the $x=2.8$ sample growth. An increase in the rate leads to crystal instability. Consequently, the growth process is more stable at the lower growth rate of 4--6 mm/h.  The crystals were cylindrical rods of 4--6 mm diameter and 50--100 mm length (Figure \ref{crystals}).  The color of the crystals was dark green indicating a correct oxidation state Cr$^{5+}$. 
Additionally, various annealing treatments were used in order to increase the crystals ductility. The temperature has to be lower than 350$^{\circ}$C or higher than 850$^{\circ}$ [\onlinecite{chem_phase}]. High temperature annealing, which requires quenching to a temperature lower than 350$^{\circ}$C, leads to immediate destruction of the crystals. Consequently, the crystals were annealed in a vacuum atmosphere of $1.2\times10^{-2}$ mbar at 300$^{\circ}$C for 6 days. Despite these precautions, Ba$_{0.3}$Sr$_{2.7}$Cr$_{2}$O$_{8}$ ($x = 2.7$) crystals were not stable. Growth conditions to obtain single crystals for $x<2.8$ are still not known.\par
The crystals for $x=2.9$ and $x=2.8$ were characterized by magnetization, neutron diffraction and inelastic neutron-scattering measurements.  Magnetic susceptibility measurements were performed in a Magnetic Property Measurement System (MPMS, Quantum Design Inc.) using a superconducting quantum interference device in magnetic fields of 0.01 and 0.1 T applied perpendicular and parallel to the $c$ axes. The size of the single crystal of Ba$_{0.2}$Sr$_{2.8}$Cr$_{2}$O$_{8}$ were not large enough to perform inelastic neutron scattering. \par
The single crystal inelastic neutron scattering measurements were performed on the cold-neutron triple-axis spectrometer V-2 FLEXX at HZB on a Ba$_{0.1}$Sr$_{2.9}$Cr$_{2}$O$_{8}$ ($x = 2.9$) crystal with mass of 4 g. The crystal was aligned in the scattering plane $(h, h, l)_{h}$ and cooled down to 2 K. The measurements were carried out using a vertically and horizontally focusing pyrolytic graphite monochromator and horizontally focusing pyrolytic graphite analyzer along with a beryllium filter. Constant-wave vector scans were performed to map the dispersion relation in the range up to 7 meV neutron energy transfer with fixed final wave vectors of $k_{f}$ = 1.2 ${\AA}^{-1}$ and $k_{f}$ = 1.55 ${\AA}^{-1}$.  The instrumental resolution with this conditions is 0.098 meV and 0.192 meV, respectively, as extracted from the FWHM of the vanadium incoherent line.
\section{\label{sec:results}Results and discussion}
\subsection{\label{subsec:one}Crystal characterization}
Powder X-ray diffraction patterns were taken at room temperature on crushed parts of the crystals before and after annealing. No change in the lattice constants and no impurities were observed. The lattice parameters shrink linearly with decreasing Ba content in accordance with Refs [\onlinecite{grundmann_influence,grundmann_structure}]. Interestingly, there is no significant change in the distances between Cr$^{5+}$ ions within a dimer (see Table \ref{j0}). \par
\begin{figure}[htt]
\includegraphics[width=0.4\textwidth]{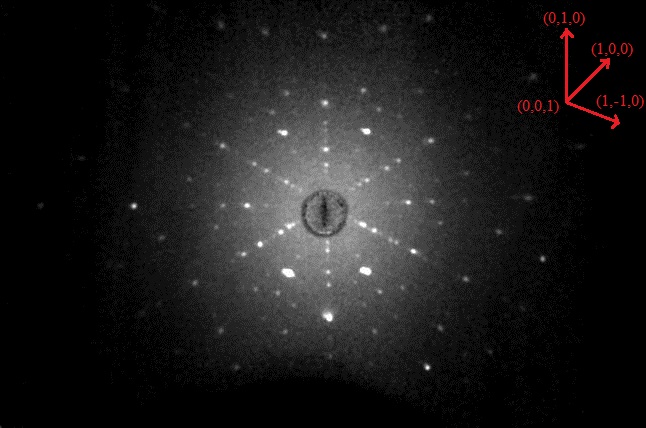}
\caption{X-ray Laue pattern of a Ba$_{0.1}$Sr$_{2.9}$Cr$_{2}$O$_{8}$ crystal showing the hexagonal symmetry at room temperature.}
\label{laue}
\end{figure}

Both crystals were aligned using Laue diffraction techniques and cut along the normal plane to the main crystal axes (Figure \ref{laue}). Energy-dispersive X-ray (EDX) analysis was used to estimate the ratio of Sr to Ba content. This value is homogeneous along the crystals and agrees well with the expected values (e.g., mixing of Ba concentration $\cong3.5\%$ vs expected value of $\cong3.33\%$), except on the surfaces of the crystals where the Ba ions tend to concentrate. Despite many advantages, the floating-zone method has a number of limitations, e.g. phase separation and incongruent melting. The first limitation can be excluded here because X-ray data show no additional phases present in the crystals. However, the physics of incongruent melting is complex because is governed by different processes such as thermal diffusion, partial melting and segregation. An incongruent melting behavior of Ba$_{3-x}$Sr$_{x}$Cr$_{2}$O$_{8}$ could result in a preferred evaporation of Ba in the center of the melt and its migration to the crystal edges.
\subsection{\label{subsec:twoone}Magnetic susceptibility}
In pure systems, spin dimerization is known to occur below 38 K in Sr$_{3}$Cr$_{2}$O$_{8}$ [\onlinecite{singh}] and 16 K in Ba$_{3}$Cr$_{2}$O$_{8}$ [\onlinecite{aczel_sr}] respectively. This manifests itself in a temperature-dependent magnetic susceptibility having a broad maximum and a drop at low temperatures, which is characteristic for systems with nonmagnetic spin-singlet ground states. \par
\begin{figure*}[htt]
\includegraphics[width=0.8\textwidth]{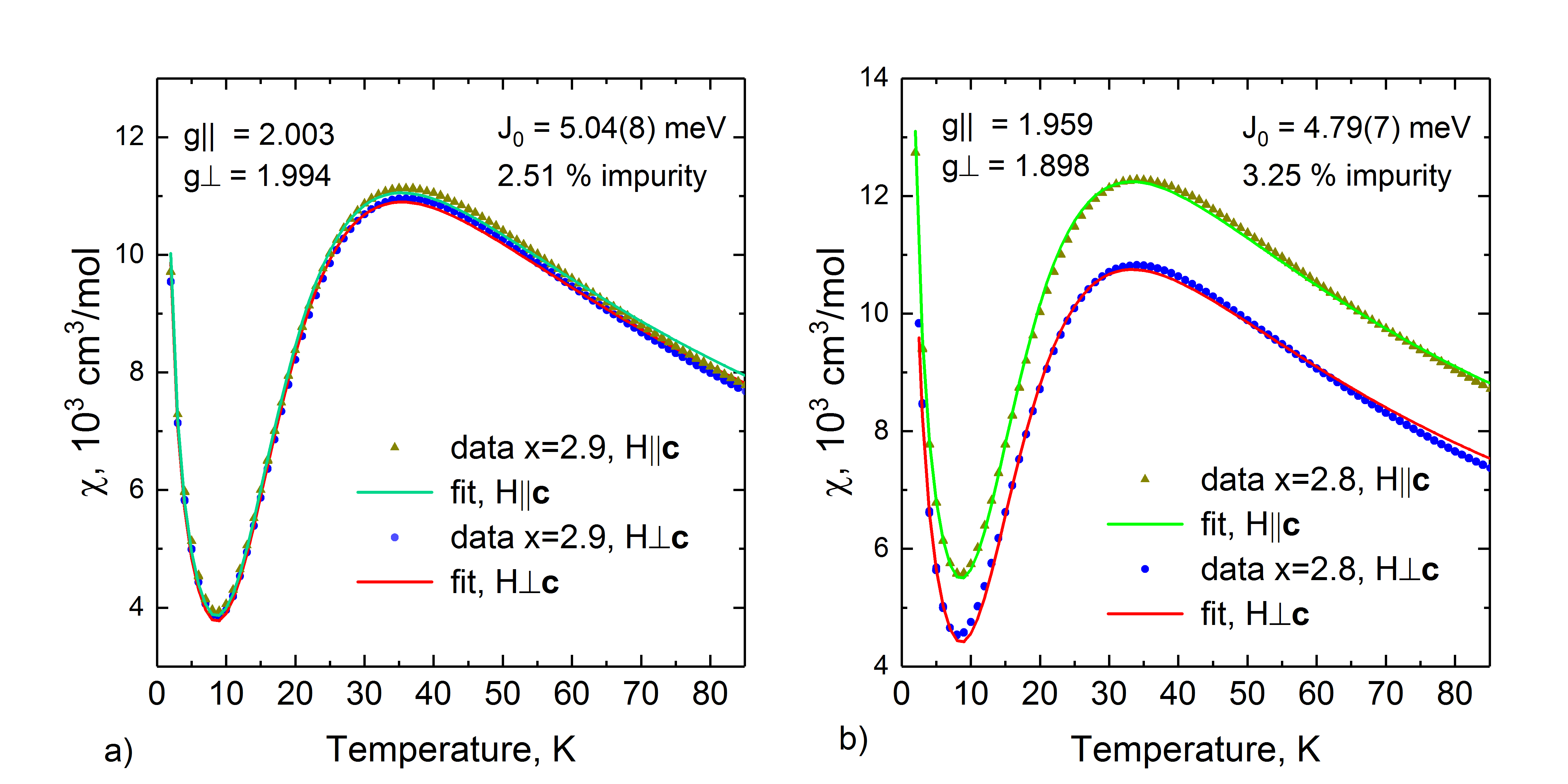}
\caption{Magnetic susceptibility data for single crystals of $a)$ Ba$_{0.1}$Sr$_{2.9}$Cr$_{2}$O$_{8}$  and $b)$ Ba$_{0.2}$Sr$_{2.8}$Cr$_{2}$O$_{8}$ with the field perpendicular and parallel to $c_{h}$, respectively.}
\label{squid} 
\end{figure*}

The magnetic behavior of the solid-solution compounds is also typical for dimerized spin systems for all other values of $x$, with an interaction constant $J_{0}$ that depends strongly on the composition [\onlinecite{grundmann_structure}]. The magnetic susceptibility measurements for  $x=2.9$ and $x=2.8$  single crystals in a 0.1 T magnetic field applied perpendicular and parallel to the $c$ axes are depicted in Figure \ref{squid}, respectively. The results are in agreement with previous measurements on  polycrystalline samples\cite{grundmann_structure} and with those for Sr$_{3}$Cr$_{2}$O$_{8}$ single crystals [\onlinecite{singh, aczel_sr, diana_mag}].
The marked increase in the magnetic susceptibility at very low temperatures suggests the presence of a fraction of paramagnetic impurities ($\cong$ 3\%, see Table \ref{j0}). Our experimental data can be well described with the Bleaney-Bowers formula (Eq.\ref{BBf}) for interacting spin$-\frac{1}{2}$-dimers, with an intradimer exchange constant $J_{0}$, and in interdimer coupling constant  $J_{e}$,
\begin{equation}
  M_d(T)=\frac{n_dg^2\mu_B^2B_{ext}}{J_e+k_BT(3+e^{\frac{J_0}{k_BT}})},
   \label{BBf}
\end{equation}
where $n_{d}$ is the density of the dimers and $g$ is the Land\'{e}  $g$-factor. The paramagnetic background can be accounted for with
\begin{equation}
\begin{split}
M_p(T)=n_pg\mu_B\frac{1}{2} & \left(2\coth\left(\frac{g\mu_BB_{ext}}{k_BT}\right)-\right.
\\
\left. \coth\left(\frac{g\mu_BB_{ext}}{2k_BT}\right)\right) \cong n_p\frac{g^2}{4}\frac{\mu_B^2B_{ext}}{k_BT},    
\end{split}
\label{BPB}
\end{equation}
where $n_p$ is the density of the corresponding uncoupled spin$-\frac{1}{2}$ ions. The sum of the above terms with $2n_{d}+n_{p}=const.$ fits the experimental data very well (see Fig. \ref{squid}), but extracting the interdimer interaction constant $J_e$ is problematic as the fit is not very sensitive to variations of $J_e$. This is also the case for Sr$_3$Cr$_2$O$_8$ [\onlinecite{diana_mag}].
The fitted intradimer interaction constants $J_0$ for our single crystal samples are presented in Table \ref{j0}, together with the $g$-factors for the different crystal orientations and magnetic impurity contents. The values for the parent compounds have been also listed for comparison. As expected, and in agreement with our previous results on polycrystalline samples [\onlinecite{grundmann_structure}], $J_0$ depends nonlinearly on the stoichiometry and decreases from $x=3$ to $x=2.9$ and $x=2.8$. The magnetic impurity content increase with increasing Ba content, i.e., consequently, decreasing Sr content $x$. It may stem from impurities undetected by X-rays, or more likely be due to intrinsic unpaired moments in the disordered crystals. \par
\begin{table}
\caption{Intradimer constant $J_0$,  Land\'{e} $g$-factors for different crystal orientations, and magnetic impurity content $n_{p}$ as extracted from DC-magnetic-susceptibility measurements}
\begin{ruledtabular}
\label{j0}
\begin{tabular}{ccccc}
                & x=3 [\onlinecite{singh}]& x=2.9 & x=2.8 & x=0 [\onlinecite{aczel_ba}] \\ \hline
\multicolumn{1}{c}{$g_{\bot}$} & 1.98 & 1.994 & 1.898 & 1.99(1)\\
\multicolumn{1}{c}{$g_{\parallel}$}  & 1.98 & 2.003 & 1.959  & 1.94(1)\\ 
\multicolumn{1}{c}{$J_0$ (meV)}  & 5.34 & 5.04(8) & 4.79(7) & 2.17(2) \\ 
\multicolumn{1}{c}{Magnetic impurities, $n_{p}$} &$\cong$ 0.2\% & 2.51\% & 3.25\% & $\cong$1\% \\ 
\multicolumn{1}{c}{Cr$^{5+}$ - Cr$^{5+}$ distance (\AA) [\onlinecite{grundmann_structure}]} & $\cong$ 3.755 & $\cong$ 3.77 & $\cong$ 3.79 &  $\cong$ 3.96 \\ 
\end{tabular}
\end{ruledtabular}
\end{table}
\subsection{\label{subsec:three}Magnetic excitations}
\begin{figure*}[htt]
\includegraphics[width=0.9\textwidth]{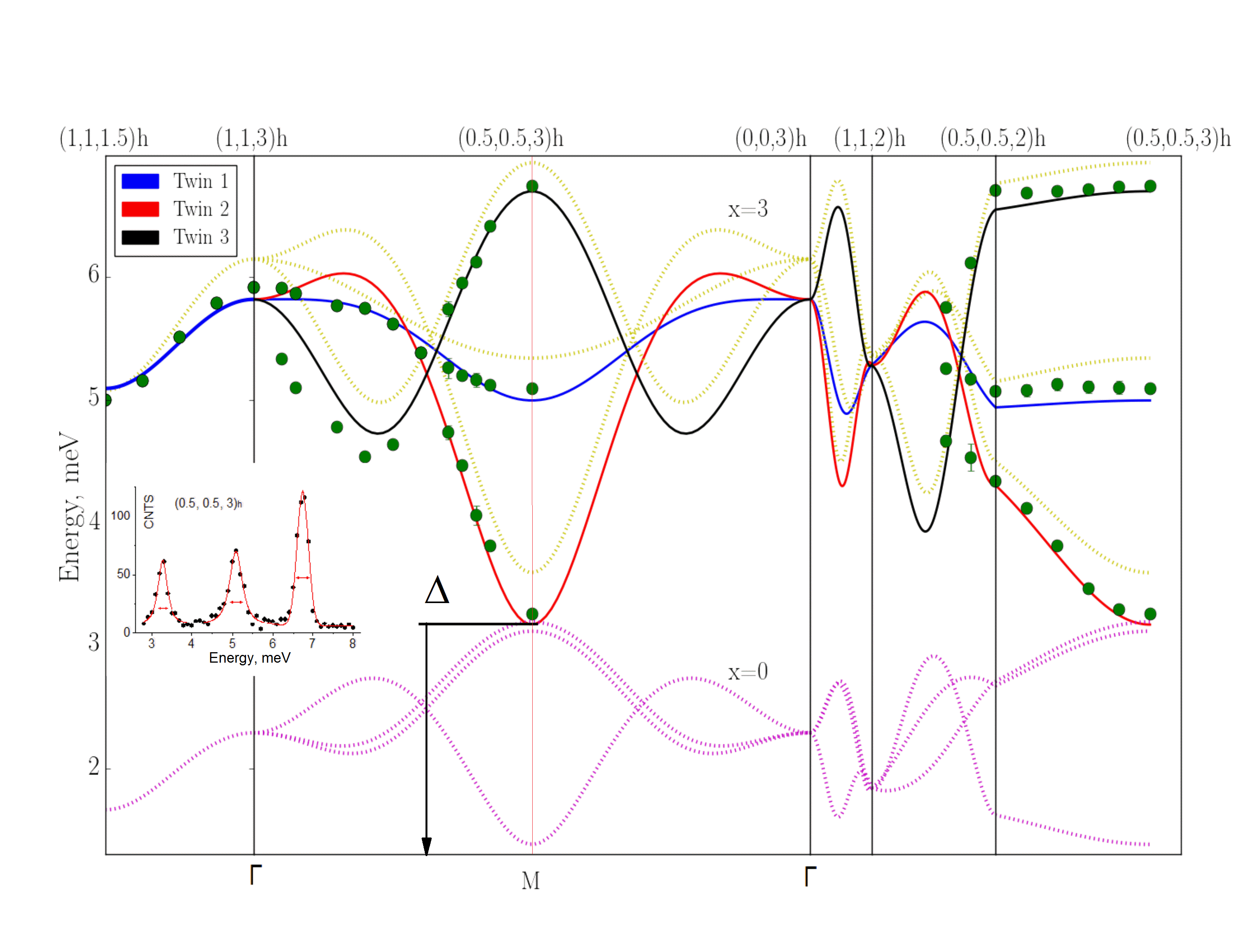}
\caption{Dispersion relation for Ba$_{0.1}$Sr$_{2.9}$Cr$_{2}$O$_{8}$ for the three monoclinic twins. The green points are the extracted peak positions from energy scans. The dispersion relation for the $x=3$ [\onlinecite{diana_mag}] and $x=0$ [\onlinecite{kofu}] compounds within RPA model are represented by yellow and violet dotted lines, respectively. The resulting RPA model parameters are listed in Table \ref{jparameters}. A single scan is shown in the inset. Horizontal lines represent the calculated instrument resolution (FWHM). }
\label{disp} 
\end{figure*}

Single-crystal inelastic neutron measurements were performed on the cold-neutron triple-axis spectrometer (TAS) V2-FLEXX at 2 K.  The dispersion relation spectrum was mapped at constant wavevector by scanning the neutron energy, and one of these energy scans is shown in the inset of Figure \ref{disp}.  In order to map out the dispersion of the singlet-to-triplet excitations, the scans were fitted to the pseudo-Voigt profile,  a linear combination of a Gaussian curve and a Lorentzian curve. \par
Figure \ref{disp} shows three modes in the dispersion. The three lines represent the three twins. These modes merge in the center of the hexagonal Brillouin zone ($\Gamma$ point).\par
We have fitted the lines by a RPA model, where the dispersion relation is defined as:
\begin{equation}
\hslash \omega \cong \sqrt{J_0^2 + J_0 \gamma(Q) },
\label{disp_rel}
\end{equation}
with $\gamma(Q)$ the Fourier sum of the interdimer interactions as described in Refs. [\onlinecite{diana_mag,kofu}]:
\begin{equation}
\gamma(Q) = \sum_{i} J(R_i)e^{-iQR_i}. 
\label{gamma}
\end{equation}
As indicated in Figure \ref{str}, the dimers (coupled Cr ions) aligned along the $c$-axis are characterized by the intradimer interaction constant $J_0$ which is of antiferromagnetic (AF) character ($J_0>$ 0). Interdimer interactions are intralayer and interlayer exchange constants and are represented by the interaction constant $J_1$ with three nearest neighbors in the adjacent layer, $J_2$ with six next neighbors in-plane, $J_3$ with six next further neighbors in the adjacent plane, and $J_4$ with three furthest neighbors in the adjacent layer. In our experiment and model, it is not possible to distinguish between $J_2$ and $J_3$, only the difference $J_2 - J_3$ can be measured. The intradimer interaction constant $J_0$ is clearly dominant. In our case we found that $J_4$ is negligible and it is not further taken into account. The low temperature structure is monoclinic with three twins which are rotated by 60$^{\circ}$ with respect to each other, and the interactions  $J_1$,  $J_2$ and $J_3$ become inequivalent. 
The interaction $J_1$ then splits in three interactions $J_1'$, $J_1''$ and $J_1'''$, $J_2$ (together with $J_3$) in two $J_2'$, two $J_2''$ and two $J_2'''$. The value of the effective interdimer interactions is $J_e =  |J_1'| + |J_1''| + |J_1'''| + 2(|J_2' - J_3'|) + 2(|J_2'' - J_3''|) + 2(|J_2''' - J_3'''|)$ [\onlinecite{diana_mag}]. \par
The resulting dispersion relation for the three monoclinic twins is illustrated in Figure \ref{disp}, where a comparison with the calculated dispersion relations for $x=2.9$, $x=3$ and $x=0$ is also shown. 
From the six values $J_1'$, $J_{2-3}' = J_2' - J_3'$ (and the corresponding "$''$" and "$'''$" constants) for $twin$ $1$, the related constants for the other twins can be found. The physical meaning of these interaction constants is described in detail in Ref. [\onlinecite{diana_mag}]. Our results are presented in Table \ref{jparameters}, along with the corresponding exchange constants for Sr$_{3}$Cr$_{2}$O$_{8}$ ($x=0$) and Ba$_{3}$Cr$_{2}$O$_{8}$ ($x=3$) from the literature [\onlinecite{diana_mag,kofu}]. The basic magnetic structure does not change but it should be noticed that $J_{2-3}'$ and $J_{2-3}'''$ have different sign in Ba$_{3}$Cr$_{2}$O$_{8}$, which indicates a tendency to ferromagnetic couplings for compounds closer to Sr$_{3}$Cr$_{2}$O$_{8}$. 
\begin{table}
\caption{Characteristic parameters for the structural transition, triplon condensation and exchange constants of Ba$_{0.1}$Sr$_{2.9}$Cr$_{2}$O$_{8}$ ($x=2.9$) as compared to corresponding values for Sr$_{3}$Cr$_{2}$O$_{8}$ ($x=3$) and Ba$_{3}$Cr$_{2}$O$_{8}$ ($x=0$)}
\begin{ruledtabular}
\begin{tabular}{{cccc}}
\label{jparameters}
 & x=3 [\onlinecite{diana_mag}]       & x=2.9     & x=0 [\onlinecite{kofu}]    \\ \hline
Structural transition  &        &  \\ 
$T_{JT}$                & 285 K  & 260 K  & 70 K   \\ \hline
Exchange Constants (meV) &        &  \\ 
$J_0$                & 5.551(9)  & 5.332(2)  & 2.38   \\ 
$J_1'$               & -0.04(1)  & -0.07(9)  & -0.15  \\ 
$J_1''$              & 0.24(1)   & 0.20(3)   & 0.08   \\ 
$J_1'''$             & 0.25(1)   & 0.24(5)   & 0.10   \\ 
$(J_2'-J_3')$          & 0.751(9)  & 0.777(5)   & 0.10   \\ 
$(J_2''-J_3'')$        & -0.543(9) & -0.528(1)  & -0.52  \\
$(J_2'''-J_3''')$      & -0.120(9) & -0.112(1) & 0.07   \\
$J_4'$               & 0.10(2)   & 0 & 0.10   \\
$J_4''$              & -0.05(1)  & 0   & 0.04   \\
$J_4'''$             & 0.04(1)   &  0   & 0.09   \\
$J_e$                & 3.6(1)    & 3.36(2)    & 1.94   \\ 
$J_e/J_0$            & 0.64(2)   & 0.63(1)   & 0.8151 \\ \hline
Triplon condensation &        &  \\
$\mu_0H_{c1}$                & 30.4 T & 28.8 T  & 12.5 T 
\end{tabular}
\end{ruledtabular}
\end{table}
The influence of disorder on the structural phase transition along with the linear change of lattice parameters were previously analyzed on polycrystalline Ba$_{3-x}$Sr$_{x}$Cr$_{2}$O$_{8}$ samples with varying stoichiometry [\onlinecite{grundmann_influence}]. Interestingly, the intradimer interaction constant calculated from the neutron powder diffraction data was reported to change nonlinearly with Sr content $x$. The peculiar behavior of $J_{0}$ can be explained by the changes in the average crystal structure. This result is commensurate with previously extracted values of $J_{0}$ from magnetization measurements [\onlinecite{grundmann_structure}]. The Jahn-Teller distortion induces an orbital ordering [\onlinecite{chapon}] and increases the intradimer interaction constant $J_{0}$ but seems to be gradually suppressed for intermediate values of $x$. A larger value of $J_{0}$ is accompanied by a stronger symmetry breaking, in other words, by increasing disorder in the system. The tuning of $J_{0}$ with varying $x$ therefore allows a direct control of the critical magnetic fields $H_{c}$ in Ba$_{3-x}$Sr$_{x}$Cr$_{2}$O$_{8}$ [\onlinecite{grundmann_tuning}].\par
The smaller lattice distances in Sr$_{3}$Cr$_{2}$O$_{8}$ than in Ba$_{3}$Cr$_{2}$O$_{8}$ lead to a larger intradimer interaction constant and spin gap. The values of the intradimer interaction constants $J_0$ are smaller for Ba$_{3}$Cr$_{2}$O$_{8}$ (2.38 meV) and Ba$_{0.1}$Sr$_{2.9}$Cr$_{2}$O$_{8}$ (5.332(2) meV) when compared with Sr$_{3}$Cr$_{2}$O$_{8}$ (5.551(9) meV). This can be easily explained by the comparably shorter atomic distances in Sr$_{3}$Cr$_{2}$O$_{8}$. The DM anisotropy term should also play an important role on the not-equal exchange interactions found. The ratio of the excitation bandwidth $J_e$ to the average mode energy, $J_e/J_0$, is larger in Sr$_{3}$Cr$_{2}$O$_{8}$ (0.64(2) meV) than for Ba$_{0.1}$Sr$_{2.9}$Cr$_{2}$O$_{8}$ (0.63(1) meV) and smaller than for Ba$_{3}$Cr$_{2}$O$_{8}$ (0.8151 meV). The resulting spin gap in Sr$_{3}$Cr$_{2}$O$_{8}$, 3.451(8) meV, is larger than in Ba$_{0.1}$Sr$_{2.9}$Cr$_{2}$O$_{8}$ (3.174 meV) and Ba$_{3}$Cr$_{2}$O$_{8}$ (1.38 meV). 
We can therefore expect that the critical magnetic field where the spin gap closes, (i.e. the onset of BEC), is smaller in Ba$_{0.1}$Sr$_{2.9}$Cr$_{2}$O$_{8}$ than in Sr$_{3}$Cr$_{2}$O$_{8}$, by roughly 8\%. From high-field magnetization measurements on polycrystalline samples at $T = 1.5$ K, [\onlinecite{grundmann_tuning}], we indeed inferred  a $\mu_0H_{c1}{\cong}$ 28.8 T for Ba$_{0.1}$Sr$_{2.9}$Cr$_{2}$O$_{8}$, as compared to 30.4 T in Sr$_{3}$Cr$_{2}$O$_{8}$.\par
The inset in Figure \ref{disp} shows one of the multiple TAS scan at $(0.5, 0.5, 3)_{h}$. The three peaks are only slightly broader than the instrumental resolution, indicating long-range exitations. Scans were also extended below the gap energy down to 1.7 meV, but no in-gap intensity was found. These two characteristics show that there is no considerable magnetic disorder caused by the random substitution of Sr by Ba. \par
The value of the intradimer interaction constant for $x = 2.9$ (5.04 meV) as obtained by magnetization measurements is somewhat lower than that from fitting the inelastic neutron-scattering data to a random phase approximation (RPA) model (5.332 meV). This may reflect a general trend, as $J_0$ values reported for $x=3$ (Sr$_3$Cr$_2$O$_8$) from magnetic susceptibility data vary from 5.34 meV [\onlinecite{singh}], 5.30 meV [\onlinecite{aczel_ba}] to 5.51 meV [\onlinecite{diana_mag}], while INS data yield a larger value, $J_0$ = 5.55 meV [\onlinecite{diana_mag}]. A similar feature has been observed also in corresponding data for $x=0$ (Ba$_3$Cr$_2$O$_8$) with $J_0 \cong 2.17(2)$ meV (magnetization) and 2.38 meV (INS) , respectively [\onlinecite{aczel_ba}]. We believe that a comparison of such data within the same measurement technique (i.e., within the Tables \ref{j0} and \ref{jparameters}) yields the correct trend, while the true values of $J_0$ may be closer to those obtained by nelastic neutron scattering data. However, both RPA and  the Bleaney-Bowers approximation are mean field approximations to the multi-body phenomena. INS gives more precise values for $J_e$ because corresponding fits are done to larger data sets involving several directions in reciprocal space. 

\section{\label{sec:four}Conclusions}
The conditions to grow single crystals of Ba$_{0.1}$Sr$_{2.9}$Cr$_{2}$O$_{8}$ and Ba$_{0.2}$Sr$_{2.8}$Cr$_{2}$O$_{8}$ by floating zone method were examined, and their structural and magnetic properties were explored. The magnetic susceptibilities for different field orientations indicate a certain anisotropy of the Land\'{e}  g-factor for magnetic fields parallel or perpendicular to the $c_{h}$-axes, respectively, which is larger in Ba$_{0.2}$Sr$_{2.8}$Cr$_{2}$O$_{8}$ than in Ba$_{0.1}$Sr$_{2.9}$Cr$_{2}$O$_{8}$.  Inelastic neutron scattering measurements of Ba$_{0.1}$Sr$_{2.9}$Cr$_{2}$O$_{8}$  confirm a change of the magnetic interaction constants and the spin gap upon a partial substitution of Sr by Ba. The observed dispersion relation is excellently reproduced by the RPA model. No sign of magnetic disorder is detected, neither in form of the appearance of in-gap intensity, nor as a noticeable energy broadening of the excitations. The intradimer interaction $J_{0}$ decreases for $x=2.9$ by about 4\% as compared to $x=3$ when chemical disorder is introduced, while the effective interdimer interaction decreases by about 7\%, and the spin gap is reduced by about 8\%. However, the $x=2.9$ compound does not show any magnetic disorder. Higher substitution levels in this family of compounds may provide a playground to study disorder effects in mixed spin-dimer systems. Unfortunately, inelastic neutron scattering experiment failed on Ba$_{0.2}$Sr$_{2.8}$Cr$_{2}$O$_{8}$ crystal due to instability, which lead to a decrease in crystal size, making these experiment impossible. Further investigations on the nuclear crystal structure are underway and will be provided in a separate publication.
\begin{acknowledgments}
We thank  Bella Lake for the possibility to realize this research, A. T. M. Nazmul Islam for the help with crystal growth and the Helmholz Zentrum Berlin for the access to neutron beamtime at the research reactor BER II. We thank Ekaterina Pomjakushina for EDX analysis. 
This work was supported by the Swiss National Science Foundation Grant No. 21-153659. 
\end{acknowledgments}

\end{document}